\documentstyle[neuchatel]{article}

\newcommand{\axion}{\hbox{\Large $a$} }
\newcommand{\Zint}{{\mbox{\sf Z\hspace{-2.7mm} Z}}}

\begin{document}
%
%
%
{
\baselineskip 16pt
\begin{titlepage}
 
{\large\baselineskip18pt
 
\hfill CERN-TH/97-368
 
\hfill CPTH-PC589.1297
 
\hfill {\tt hep-th/9712155}
 
}
 
\begin{center}
\vfill
{\Large\bf D-effects in Toroidally Compactified Type II String Theory
~\footnotemark}
\footnotetext{\large ~To appear in the proceedings of the conference
{\sl ``Quantum Aspects of Gauge Theories, Supersymmetry and Unification''}, 
held at Neuch\^atel University, Neuch\^atel, Switzerland,
18--23  September 1997.}
\end{center}
\vskip .8in
\begin{center}
{\Large Boris  Pioline}
 
\vskip .4in
 
{\large\sl Theory Division, CERN, CH-1211 Geneva 23}
 
{\sl and}
 

{\large\sl Centre de Physique Th\'eorique, Ecole Polytechnique, F-91128 Palaiseau Cedex}

 \end{center}
 
\vskip1in
 
\begin{quotation}
{\large\rm We review exact results obtained for $R^4$ couplings in maximally
supersymmetric type II string theories. These couplings offer
a privileged scene to understand the rules of semiclassical
calculus in string theory. Upon expansion in weak string coupling,
they reveal an infinite sum of non-perturbative $e^{-1/g}$ effects
that can be imputed to euclidean D-branes wrapped on cycles
of the compactification manifolds. They also shed light on the
relation between Dp-branes and D-(p-2)branes, D-strings and
$(p,q)$ strings, instanton sums and soliton loops. The latter 
interpretation takes over in $D\le 6$ in order to account
for the $e^{-1/g^2}$ effects, still mysterious
from the point of view of instanton calculus.
}
 \end{quotation}
\vfill
\noindent {\large CERN-TH/97-368, CPTH-PC589.1297}
 
\noindent {\large December, 1997}
\vskip12pt
\end{titlepage}
}
\vfill\eject
\addtocounter{footnote}{-1}
 
%
%

\def\titleline{
D-effects in Toroidally Compactified Type II String Theory
}
\def\authors{
Boris Pioline\1ad \2ad
}
\def\addresses{
\1ad
Theory Division, CERN, CH-1211, Geneva 23\\
\2ad
Centre de Physique Th\'eorique, Ecole Polytechnique, F-91128 Palaiseau Cedex
}
\def\abstracttext{
We review exact results obtained for $R^4$ couplings in maximally
supersymmetric type II string theories. These couplings offer
a privileged scene to understand the rules of semiclassical
calculus in string theory. Upon expansion in weak string coupling,
they reveal an infinite sum of non-perturbative $e^{-1/g}$ effects
that can be imputed to euclidean D-branes wrapped on cycles
of the compactification manifolds. They also shed light on the
relation between Dp-branes and D-(p-2)branes, D-strings and
$(p,q)$ strings, instanton sums and soliton loops. The latter 
interpretation takes over in $D\le 6$ in order to account
for the $e^{-1/g^2}$ effects, still mysterious
from the point of view of instanton calculus.

}
\makefront
\section{Introduction}
Extending string theory beyond the perturbative regime has become
an important challenge over the last two years. Whereas string
theory was originally formulated as a perturbative genus expansion,
string duality has allowed the determination of a number of exact
non-perturbative results, starting with the vector multiplet geometry
in $N=2$ four-dimensional string theories. In that case, the metric
is given exactly at tree-level on the type II side, and the $e^{-1/\alpha'}$
world-sheet instanton corrections translate into $e^{-1/g^2}$ space-time
instantons on the Heterotic side. On the other hand, the large order 
behaviour of the string perturbation series allows much stronger
$e^{-1/g}$ non-perturbative effects, a typical feature of the old matrix 
models \cite{shenker}. While such instanton effects are absent 
from ordinary gauge theories, they are naturally accommodated 
in string theory as the D-branes of type II theories, or their 
Wick rotated analogues. Their contributions to various 
couplings may be determined by duality, and one may hope to
learn the rules for including and weighting them
from a close scrutiny of these results.
Although less ambitious, this approach aims at extending
string theory into the semi-classical regime. This is only
meaningful if the perturbative corrections can be brought under control, 
an achievement of supersymmetry in many cases.

Due to their classical action action 
$e^{-1/\sqrt{\rm Im S}}$ being non-holomorphic in the chiral
dilaton field $S$, D-effects cannot correct the vector multiplet geometry.
This restriction however does not apply to the hypermultiplet geometry,
and this is where this type of contribution was first investigated
\cite{bbs}. In the case of a single hypermultiplet, the symmetry
restrictions on the hyperk\"ahler metric are strong enough to determine
it uniquely \cite{ov}. It then exhibits an infinite sum of non-perturbative
effects of order $e^{-n{\cal A}/g}$, where ${\cal A}$ is the area of
the three-cycle parametrized by the hypermultiplet, and $n$ the
summation integer. These effects can be interpreted as euclidean D2-branes
wrapped $n$ times around the three-cycle, and altogether resolve the 
the logarithmic conifold singularity of the metric. This is dual to the
resolution of the conifold singularity on the vector moduli space,
which occurs via loops of massless solitons \cite{strominger}.

Significant progress in the understanding of the rules of instanton 
calculus has been achieved by examining simpler settings with more 
supersymmetries, namely toroidal compactifications of type II strings.
U-duality often allows to completely determine amplitudes, which due
to their supersymmetric structure can receive corrections 
from BPS saturated instanton configurations only.
Upon expansion in the weak coupling regime, they reveal infinite sums
of non-perturbative effects, corresponding to D-branes wrapped on
various cycles of the compactification torus. After recalling in
Section 2 some background on instanton corrections in string theory,
we shall review in Section 3 the results obtained on the exact $R^4$ 
couplings in $N=8$ vacua
\footnote{SUSY charges are counted in four-dimensional units.}
and their interpretation. The reader is refered
to Ref.\ \cite{apt} for a detailed discussion on instanton corrections
in Type II string theory compactified on a curved manifold $K_3$
corresponding to the second part of my talk at this workshop.
Related work in the context of type I string theory
appeared in Ref.\ \cite{bfkov}.

\section{Instanton corrections and BPS-saturated amplitudes.}
Instantons of ordinary Quantum Field Theories are saddle points of
the Euclidean fundamental action $S$. They usually come in continuous
families, and the integration measure $\mu$ for their collectives coordinates
$\phi$
can be obtained by a change of variables from the canonical fields.
Part of the collective coordinates are generated by
the symmetries broken by the instanton background
and which are restored upon integration over $\phi$.
The Gaussian fluctuations around the instanton background give rise
to the fluctuation determinant $\det Q$, 
and the correlation functions take the
symbolic form
\begin{equation}
\langle \hskip 1cm \rangle = 
\sum_{\mbox{sectors}} \int d\mu 
\langle \hskip 1cm \rangle_I\frac{1}{\sqrt{\det Q}} e^{-S_I}
\end{equation}
In the presence of supersymmetry, the bosonic and fermionic
oscillators cancel in the determinant, leaving the zero-modes only. The SUSY
generators broken by the instanton background generate
fermionic zero-modes that have to be saturated by a sufficient
number of vertex insertions. This restricts the terms in
the effective action which can receive corrections. 
When supersymmetry is extended, instantons can break SUSY only partially,
and accordingly, terms in the effective action which are related to
fermionic terms with a low number of fermions, can receive corrections
from a restricted set of instantons. These are the so called 
{\it BPS saturated} couplings. The $1/2$-BPS saturated couplings
include the $R^4$ couplings
in $N=8$, $R^2,F^4$ couplings in $N=4$, the kinetic terms of
$N=2$ (including the hypermultiplet metric), and the D-terms
of the scalar potential in $N=1$.

Unfortunately, string theory so far does not come with a fundamental action,
and the rules of instanton calculus have to be inferred rather than derived.
It is however still possible to use the supergravity low-energy description of
string theory to determine the possible $1/2$-BPS instantons. These include
the Neveu-Schwartz (NS) 5-brane, and the type II D-branes. In order to 
yield finite action saddle points, such objects have to be wrapped on
cycles of the compactification manifold. This in particular means
that $1/2$-BPS saturated couplings in ten-dimensional heterotic string, 
type IIA and type I are exact in perturbation theory, while the type IIB
couplings can receive corrections from the D-instantons already in 
ten dimensions. This is the case we now turn to.

\section{$R^4$ couplings in maximally supersymmetric theories} 
The $R^4$ coupling in ten-dimensional type IIB theory offers the most
simple setting to investigate instanton corrections. It is
given in the Einstein frame
by a function $f_{10}^B(\tau,\bar\tau)$ of the only
scalar $\tau=\axion+i~e^{-\phi}$ (where $e^{\phi}=g$ is the string
coupling and $\axion$ the Ramond scalar):
\begin{equation}
\label{1}
S_{R^4}= \int d^{10}x\sqrt{-g_{E}}
~f_{10}^B(\tau,\bar\tau)~
(t_8t_8 + \frac{1}{8} \epsilon_{10}\epsilon_{10}) R^4
\end{equation}
which can be evaluated at tree-level and one-loop order:
\begin{equation}
\label{2}
f_{10}^B(\tau,\bar\tau)= 
e^{\phi/2}~\left( 2\zeta(3) e^{-2\phi} + \frac{2\pi^2}{3} \right)
\end{equation}
As it stands, $f_{10}^B$ is not invariant under the type IIB
$Sl(2,\Zint)_B$ symmetry, under which $\tau$ transforms as
a modular parameter. It was realized in Ref.\ \cite{gg} that
one could supplemented it by an infinite sum of non-perturbative effects
which restore the $Sl(2,\Zint)_B$ invariance:
\begin{equation}
\label{3}
f_{10}^B
= 2\zeta(3) e^{-3\phi/2} +  \frac{2\pi^2}{3} e^{\phi/2}
+ 4\pi e^{-\phi/2} \sum_{m\ne 0}  \sum_{n\ne 0}
\left|\frac{m}{n}\right| K_1\left(2\pi e^{-\phi} |mn|\right) e^{2\pi i m n 
\axion}
=\sum_{(m,n)\ne 0} \left( \frac{\tau_2}{|m+n\tau|^2} \right)^{3/2}
\end{equation}
The above equality follows by Poisson resummation on the integer $m$,
after separating the contributions $(m\ne 0,n=0)$. $\zeta(3)$ is
Apery's transcendental number.
Using the saddle point approximation $K_1(z)=\sqrt{\pi\over 2 z} e^{-z}
\left( 1+ O(1/z)\right)$ of the Bessel function, the double sum
approximates to 
\begin{equation}
\label{4}
2\pi \sum_{m\ne 0}  \sum_{n\ne 0} { |mn|^{1/2} \over n^2}
e^{-2\pi |mn| (e^{-\phi} \pm i a)} \left( 1+ O(e^{\phi}) \right) \nonumber
\end{equation}
which has the right form to be interpreted as a sum of D-instantons
effects with action $S_{cl}=a + i~e^{-\phi}$. The product of integers
$N=mn$ can be interpreted as the charge of the D-instantons, or alternatively
the number of elementary D-instantons put together, while the
measure $\mu(N)=\sum_{n|N}{1\over n^2}$ should be computable from
a Matrix model analysis \cite{gg2}. $O(e^{-\phi})$ corrections correspond 
to perturbative corrections around the instanton background.

Although the conjecture in Eq.\ (\ref{3}) has not been rigorously proved
due to the lack of mathematical knowledge about Eisenstein series, strong
evidence has already been given \cite{gv,apt,berkovits}. 
It has also been extended to other type IIB ten-dimensional 1/2-BPS couplings
in Refs. \cite{otherbps}. We now want to  extend
this conjecture to type IIB theory compactified on a N-dimensional
torus $T^N$. That these lower-dimensional
results have a sensible D-brane expansion will strengthen the evidence
for Eq.\ (\ref{3}). Under compactification on a circle, no further
instantons appear in the type IIB theory
(in agreement with the fact that the U-duality group
is still $Sl(2,\Zint)$), so that the $R^4$ coupling
is simply obtained from Eq.\ (\ref{3}) by multiplying by the length
of the circle. This is easily translated on the type IIA
side by T-duality:
\begin{equation}
\label{5}
f_{9}^A=2\zeta(3) r_A e^{-2\phi} +  
         \frac{2\pi^2}{3 r_A} + 0
+~ 4\pi e^{-\phi} \sum_{m\ne 0} \sum_{n\ne 0}
\left|\frac{m}{n}\right| K_1\left(2\pi r_A e^{-\phi} |mn|\right) 
          e^{2\pi i m n {\cal A}}
\end{equation}
reproducing the tree-level and  one-loop contributions together
with the contributions of D0-brane instantons wrapped on the circle,
with action $S_{cl}={\cal A} + i~r_A e^{-\phi}$, where ${\cal A}$ is
the Wilson line of the Ramond one-form on the circle
\footnote{Here and henceforth, curl letters denote Ramond fields.}. 
Upon decompactification
to ten dimensions, all non-perturbative corrections together with the
(degenerate) one-loop term vanish, and it follows that the $R^4$
coupling is exact at one-loop in ten dimensions
\footnote{There is also a one-loop $R^4$ coupling but
with a different kinematical structure. The latter would 
also subsist in the M-theory decompactification limit.}. 

We now would like to determine the contributions from higher branes
to $R^4$ couplings. This can be achieved by arranging the moduli
into a symmetric coset representative $M$ transforming in the
adjoint of the $E_{11-D}(\Zint)$ U-duality group, 
and postulating that the result is given by the Eisenstein series
\begin{equation}
f_D (M)= 
\sum_{m\ne 0} \left[ m^t M m \right]^{-{3\over 2}}
\end{equation}
This was carried out in Ref.\ \cite{kp} for compactifications on 
$T^2$ and $T^3$, and the Eisenstein series was shown to reproduce
the expected $(p,q)$ string instanton corrections together with
the perturbative terms. It however becomes rather impractical
for compactifications to lower dimensions, and we instead
follow another route \cite{pk}.
The tree-level, (degenerate) one-loop and D0-brane contributions
can actually be written for any toroidal compactification,
by simply replacing the circle of radius $n r_A$ by any cycle 
of radius $\sqrt{n^i g_{ij} n^j}$, where $g_{ij}$ denotes the metric on the
$N$-torus with volume $V$:
\begin{equation}
\label{6}
f^A_{D0}=2\zeta(3) V e^{-2\phi} + 2V \sum_{n^i\ne 0} \frac{1}{n^i g_{ij} n^j}
+4\pi V \tau_2  \sum_{m\ne 0}  \sum_{n^i\ne 0}
\frac{ |m| }
{\sqrt{n^i g_{ij} n^j}} K_1\left(2\pi|m|e^{-\phi}\sqrt{n^i g_{ij} n^j}\right)
e^{2\pi i m n^i {\cal A}_i}
\end{equation}
This expansion can actually be obtained from a 
(formaly divergent) one-loop computation in eleven-dimensional supergravity
\cite{ggv}
\begin{equation}
\label{7}
f^A_{D0} = 2\pi V_{11} \int_{0}^{\infty} \frac{dt}{t^{5/2}}
\sum_{n^I\ne 0} e^{-\frac{\pi}{t} {\cal M}^2} 
= V_{11} \sum_{n^I\ne 0} \left({\cal M}^2\right)^{-3/2}, \quad
{\cal M}^2 = n^I g_{IJ} n^J,
\end{equation}
where $g_{IJ}$ now denotes the volume $V_{11}$ metric on the $N+1$-torus.
The equality of Eqs. (\ref{6}) and (\ref{7}) follow by Poisson resummation
on the integer $n^{11}$.
This is in agreement with the fact that the D0-branes are the 
Kaluza-Klein modes of the supergravity multiplet
upon reduction to ten dimensions \cite{witten1}.

Equation (\ref{7}) offers a very good starting point for determining
the contributions of higher branes to $R^4$ couplings in toroidally
compactified type II theories. Indeed, all D$p$-branes can be reached
from the D0-branes (or from the D-instantons, for that matter) by a sequence
of T-dualities. Indeed, decomposing the metric on the torus as
\begin{equation}
\label{8}
 ds^2=R^2 (dx^{1} + A_a dx^a)^2 + dx^a g_{ab} dx^b\ ,
\end{equation}
where the indices $a,b$ run from 1 to $N-1$,
and applying a T-duality
\begin{equation}
\label{9}
R\leftrightarrow 1/R,\quad
A_a \leftrightarrow B_{1a},\quad
{\cal A}_a \leftrightarrow {\cal B}_{1a},\quad
{\cal A}_1 \leftrightarrow a,\quad
B_{ab} \leftrightarrow B_{ab} -A_a B_b + B_a A_b,\quad
e^{-2\phi} R = {\rm const.}, 
\end{equation}
the classical action of a D0-brane 
\begin{equation}
\label{10}
S_{D0} = e^{-\phi} \sqrt{n^i g_{ij} n^j} + i~n^i {\cal A}_i
\end{equation}
turns into
\begin{equation}
\label{11}
S_{D1} = e^{-\phi} \sqrt{ \left( n + \frac{1}{2} n^{ij} B_{ij} \right)^2
         + \frac{1}{2} n^{ij}~g_{ik} g_{jl}~n^{kl} }
+ i~ \left(n \axion + \frac{1}{2} n^{ij} \tilde{\cal B}_{ij} \right)\ .
\end{equation}
where we reinterpreted $n^1$ as a singlet $n$, and $n^{a}$ as the component
of a two-form $n^{1a}$. The integers $n^{ab}$ in Eq.\ (\ref{11})
do not have a counterpart in Eq. (\ref{10}), but their presence is
inferred from the requirement of $Sl(N,\Zint)$ symmetry. 
$\tilde{\cal B}$ coincides with the Ramond two-form 
up to a linear shift in $B$ \cite{pk}. Equation
(\ref{11}) can be interpreted as the classical action of a
D-string wrapped on a two cycle of the torus.  
Indeed, evaluating the Born-Infeld action
\begin{equation}
\label{12}
S_{BI} = \int \frac{1}{g} \sqrt{\det(\hat G + \hat B  + F)}
+ i e^{\hat B + F} \wedge{\cal R}\ , 
\end{equation}
on a configuration
\begin{equation}
\label{13}
X^i(\sigma^\alpha) = N^i_\alpha \sigma^\alpha, \quad
n^{ij}=\epsilon^{\alpha\beta} N^i_\alpha N^j_\beta, \quad
F_{\alpha\beta}=\epsilon_{\alpha\beta} n
\end{equation}
precisely reproduces Eq. (\ref{11}).
The action for $n^{ij}=0$ reduces to the action of $n$ type IIB
D-instantons. The integer $n$, corresponding to the flux of the
U(1) gauge field on the D-string, therefore describe the number
of D-instantons bound to the string. The contribution of D-strings 
and D-instantons can therefore be summarized in the following
series:
\begin{equation}
\label{14}
f_{D1}^B=
4\pi V e^{-\phi} \sum_{m\ne 0} \sum_{(n, n^{ij})\ne 0} 
{ |m| \over \sqrt{ \det( \hat G + \hat B + F ) }}
K_1 \left( 2\pi |m| e^{-\phi} \int \sqrt{\det(\hat G + \hat B +F)} \right)
e^{2\pi i m \int \hat{\tilde{\cal B}} + a F}
\end{equation}
While this result is by construction T-duality invariant when $D\ge 7$
(in lower dimensions, the D-3 brane has to be included), it is not
guaranteed to be U-duality invariant. In order to check
the invariance under $Sl(2,\Zint)_B$, which together with
the T-duality $SO(N,N,\Zint)$ generates the full U-duality
group, it is convenient to perform a Poisson resummation on the
integer $n\rightarrow p$ (this amounts to going to the 
$\theta$ vacuum in the world-sheet gauge theory)
and rename the integer $m\rightarrow q$.
Under this operation, the Bessel function $K_1(z)$ turns into
$K_{1/2}(z)=\sqrt{\pi\over 2z} e^{-z}$, and the above expression
becomes
\begin{equation}
\label{15}
f_{D1}^B=4\pi V \sum_{l\ne 0} \sum_{p \wedge q =1}
\sum_{n^{ij}}
{ e^{-2\pi l |p + q \tau| \sqrt { n^{ij}~g_{ik}g_{jl}~n^{kl} }
    +2\pi i l n^{ij} ( q {\cal B}_{ij} - p B_{ij} ) }
\over
 \sqrt { n^{ij}~g_{ik}g_{jl}~n^{kl} } }
\end{equation}
This sum can now be interpreted as the contribution of
$(p,q)$ strings to the four-graviton amplitude. Indeed,
the term with $(p,q)=(1,0)$ corresponds to the 
one-loop world-sheet instantons 
on the fundamental string, and the other terms are
obtained by replacing the string tension  and the NS two-form by
\begin{equation}
\label{16}
\alpha' \rightarrow {\alpha' \over |p+q\tau|}, \quad
B_{ij} \rightarrow p B_{ij} - q {\cal B}_{ij}
\end{equation}
as appropriate for a $(p,q)$ string. Note that as expected,
only strings with $p$ and $q$ coprime contribute. This is a
rather dramatic change of perspective, since what was previously
interpreted as a sum of instanton effects now is now
described as a soliton sum on the BPS excitations of the
$(p,q)$ string. 

This sequence of T-duality can be pursued in order to obtain
the contributions of higher branes:
\begin{eqnarray}
\label{17}
f_{D2}^A=&e^{-\phi} \left[
\left( n^i + \frac{1}{2} n^{ijk} B_{jk} \right) g_{il}
\left( n^l + \frac{1}{2} n^{lmn} B_{mn} \right)
+ \frac{1}{6} n^{ijk}~g_{il}g_{jm}g_{kn}~n^{lmn} \right]^{1/2}\nonumber\\
&+ i~ \left( n^i {\cal A}_i + \frac{1}{6} n^{ijk} {\cal C}_{ijk} \right)
\end{eqnarray}
\begin{eqnarray}
\label{18}
f_{D3}^B=e^{-\phi} &\left[ 
\left(n + \frac{1}{2}n^{ij} B_{ij} +\frac{1}{8} n^{ijkl} B_{ij} B_{kl} 
\right)^2
+ \frac{1}{2} \left(n^{ij} + \frac{1}{2} n^{ijkl} B_{kl} \right)
g_{im}g_{jn}  \left(n^{mn} + \frac{1}{2} n^{mnpq} B_{pq} \right) \right. 
\nonumber\\
&\left.  
+ \frac{1}{24} n^{ijkl} g_{im}g_{jn}g_{kp}g_{lq} n^{mnpq}  \right]^{1/2}
+ i~\left( n \axion + \frac{1}{2}
 n^{ij} \tilde{\cal B}_{ij} 
+ \frac{1}{24} n^{ijkl} \tilde{{\cal D}}_{ijkl}  \right)
\end{eqnarray}
which can again be interpreted as the Born-Infeld action for D2- (resp. D3-)
branes wrapped on three-cycles $n^{ijk}$ (resp. $n^{ijkl}$), with 
fluxes $n^i$ (resp. $n^{ij}$). The integer $n$ is now interpreted
as the gauge instanton number 
${1\over 8}\epsilon^{\alpha\beta\gamma\delta}F_{\alpha\beta}F_{\gamma\delta}$
on the D3-brane world-volume. By construction, these actions are again
T-duality invariant for $D\ge 6$ (resp. $D\ge 5$). In fact, one can
see the expression under the square root of Eq. (\ref{17}) and 
(\ref{18}) as the norm of the two spinors of $SO(N,N)$ induced
by the coset representative $g+B$ of $SO(N,N)/(SO(N)\times SO(N))$, 
after decomposition in terms of $Sl(N)$
irreducible representations \cite{opr}. However, U-duality is not
guaranteed, and in fact not fulfilled. Indeed, going to the
representation of Eq.\ (\ref{7}), we find that the D2-brane sum
translates into 
\begin{eqnarray}
\label{19}
{\cal M}^2 = &R_{11}^2 \left( n^{11} + {\cal A}_i n^i + \frac{1}{6} n^{ijk} 
\tilde{\cal C}_{ijk} \right)^2 
+\left(n^i + \frac{1}{2} n^{ijk} B_{jk} \right)\frac{g_{il}}{R_{11} }
\left(n^l + \frac{1}{2} n^{lmn} B_{mn} \right) \nonumber\\
&+ \frac{R_{11}^2}{6}
n^{ijk}~\frac{g_{il}g_{jm}g_{kn}}{R_{11}^3} ~n^{lmn} 
\end{eqnarray}
so that $Sl(11-D,\Zint)$ invariance with respect to the metric
$ds_{11}^2 = R_{11}^2 (dx^{11} + {\cal A}_i dx^i)^2 + \frac{1}{R^{11}} 
dx^i g_{ij} dx^j$
can be only be recovered by 
extending the three-form $n^{ijk}$ into
a four-form $n^{IJKL}$:
\begin{equation}
\label{20}
{\cal M}^2 =
\left( n^I + \frac{1}{6} n^{IJKL} {\cal C}^{(11)}_{JKL} \right)
g_{IM} \left( n^M + \frac{1}{6} n^{MNPQ} {\cal C}^{(11)}_{NPQ} \right)
+ \frac{1}{24} n^{IJKL} g_{IM}~g_{JN}~g_{KP}~g_{LQ}~n^{MNPQ}\ .
\end{equation}
Whereas the D-brane charges $n^{i},n^{ijk}$ contribute $O(1/g^2)$ to 
${\cal M}^2$ (and therefore $e^{-1/g}$ after Poisson resummation
on $n^{11}$), the extra charge $n^{ijkl}$, starting to occur for $D\le 6$, 
contributes $O(1/g^4)$ to ${\cal M}^2$. It therefore belongs
to another T-duality multiplet,  and generates 
$e^{-1/g^2}$ non-perturbative effects in the corresponding 
Eisenstein series.  Such contributions are usually imputed to
the NS fivebrane, but the latter cannot generate instanton
effects when $D> 4$. The situation is not any better on the type
IIB side, where $Sl(2,\Zint)_B$ invariance requires the introduction
of an extra four-form $m^{ijkl}$ transforming as a doublet with
the D3-brane wrapping number $n^{ijkl}$.

This puzzle can be resolved
\footnote{Another way out may be that the $R^4$ couplings are not
given by Eisenstein series for $D\le 6$. An indication that this may
be the case is the fact that the $D=6$ $SO(5,5,\Zint)$ Eisenstein
series has so far resisted to our many attempts to extract the
perturbative contributions.}
 by noting that the mass formula
(\ref{20}) is nothing but the mass formula for BPS strings
of M-theory \cite{opr}, corresponding to membranes
and fivebranes wrapped on a one-cycle or a four-cycle of the
compactification torus $T^{N+1}$. These states are identified
with the states of the momentum multiplet of M-theory in the Discrete
Light-Cone Quantization \cite{egkr}, upon unwinding them from 
the light-cone circle. The $R^4$ coupling is therefore given
by the sum of the one-loop amplitudes of all BPS strings
obtained by reduction of the M-theory extended states on 
$T^{N+1}$. This is a rather plausible result, since after all
our strategy was to covariantize the fundamental string contribution
under the U-duality group.

\vskip .5cm
{\bf Acknowledgements} It is a pleasure to thank the organizers of the
Neuch\^atel workshop for providing a stimulating atmosphere, for
allowing me to give a talk and for financial support. This work
is also supported by the EEC under the TMR contract ERBFMRX-CT96-0090.
I am grateful to I. Antoniadis, E. Kiritsis,
N. Obers, and T. Taylor for an enjoyable collaboration, and
to A. Kehagias, H. Partouche and E. Rabinovici 
for helpful discussions.


\begin{thebibliography}{77}

\bibitem{shenker} S.H. Shenker, ``The Strength of Non-Perturbative Effects
in String Theory,'' presented at the Carg\`ese Workshop on Random Surfaces,
Quantum Gravity and Strings, Carg\`ese, France, May 1990, preprint
RU-90-47 (1990).

\bibitem{bbs} K. Becker, M. Becker and A. Strominger, 
Nucl. Phys.  {\bf B456} (1995) 130, hep-th/9507158.

\bibitem{ov} H. Ooguri and C. Vafa, Phys. Rev. Lett. {\bf 77} (1996) 3296,
hep-th/9608079.

\bibitem{strominger}
A. Strominger, Nucl. Phys. {\bf B451} (1995) 96, hep-th/9504090.

\bibitem{apt} I. Antoniadis, B. Pioline and T. Taylor, 
to appear in Nucl. Phys. B, hep-th/9707222.

\bibitem{bfkov} C. Bachas, C. Fabre, E. Kiritsis, N. Obers and P.
Vanhove, hep-th/9707126.

\bibitem{gg} M.B. Green and M. Gutperle,  Phys. Lett. {\bf B398}
(1997) 69, hep-th/9701093.

\bibitem{gg2}  M.B. Green and M. Gutperle, hep-th/9711107.

\bibitem{gv} M.B. Green and  P. Vanhove,
Phys. Lett. {\bf B408} (1997) 122, hep-th/9704145 hep-th/9704145.

\bibitem{berkovits} N. Berkovits, hep-th/9709116.

\bibitem{otherbps} 
J.  Russo and A.Tseytlin, hep-th/9707134; 
A. Kehagias and H. Partouche, hep-th/9710023;
M.B. Green, M. Gutperle, and H. Kwon, hep-th/9710151.

\bibitem{ggv} M.B. Green, M. Gutperle and  P. Vanhove, 
Phys. Lett. {\bf B409} (1997) 177, hep-th/9706175

\bibitem{kp} E. Kiritsis and B. Pioline,
to appear in Nucl. Phys. B., hep-th/9707018.

\bibitem{pk} B. Pioline and E. Kiritsis, to appear in 
Phys. Lett. B., hep-th/9710078.

\bibitem{witten1} E. Witten, Nucl. Phys. {\bf B443} (1995) 85, hep-th/9503124




\bibitem{opr} N. Obers, B. Pioline and E. Rabinovici, hep-th/9712084.


\bibitem{egkr} S. Elitzur, A. Giveon, D. Kutasov and E. Rabinovici,
hep-th/9707217.

\end{thebibliography}
\end{document}